\documentclass[twocolumn,superscriptaddress,showpacs,amsmath,prl]{revtex4-1}
\usepackage{amsmath}
\usepackage{mathtools} 
\usepackage{graphicx}
\usepackage{dcolumn}
\usepackage{bm}
\usepackage{hyperref}
\hypersetup{
	bookmarks=true,
	colorlinks=true,
	urlcolor=blue,
	citecolor=blue,
	linkcolor=blue, 
	pdftex,
	linktocpage=true, 
	linktoc=all,     
	hyperindex=true
}
\usepackage[mathlines]{lineno}

\usepackage[alsoload=hep]{siunitx} 
\usepackage{comment} 
\usepackage{braket} 
\usepackage{xcolor}

\usepackage{newfloat}
\DeclareFloatingEnvironment[name={Supplementary Figure},fileext=lsf,listname={List of Supplementary Figures}]{suppfigure}

\usepackage[normalem]{ulem}
\newcommand\redout{\bgroup\markoverwith
	{\textcolor{red}{\rule[.5ex]{2pt}{0.4pt}}}\ULon}

\begin{document}


\title{Roton Excitations in an Oblate Dipolar Quantum Gas}

\author{J.-N. Schmidt}%
\thanks{These authors contributed equally to this work.}
\affiliation{5. Physikalisches Institut and Center for Integrated Quantum Science and Technology, Universität Stuttgart, Pfaffenwaldring 57, 70569 Stuttgart, Germany
}%
\author{J. Hertkorn}%
\thanks{These authors contributed equally to this work.}
\affiliation{5. Physikalisches Institut and Center for Integrated Quantum Science and Technology, Universität Stuttgart, Pfaffenwaldring 57, 70569 Stuttgart, Germany
}%
\author{M. Guo}%
\affiliation{5. Physikalisches Institut and Center for Integrated Quantum Science and Technology, Universität Stuttgart, Pfaffenwaldring 57, 70569 Stuttgart, Germany
}%
\author{F. B\"ottcher}%
\affiliation{5. Physikalisches Institut and Center for Integrated Quantum Science and Technology, Universität Stuttgart, Pfaffenwaldring 57, 70569 Stuttgart, Germany
}%
\author{M. Schmidt}%
\affiliation{5. Physikalisches Institut and Center for Integrated Quantum Science and Technology, Universität Stuttgart, Pfaffenwaldring 57, 70569 Stuttgart, Germany
}%
\author{K.S.H. Ng}%
\affiliation{5. Physikalisches Institut and Center for Integrated Quantum Science and Technology, Universität Stuttgart, Pfaffenwaldring 57, 70569 Stuttgart, Germany
}%
\author{S.D. Graham}%
\affiliation{5. Physikalisches Institut and Center for Integrated Quantum Science and Technology, Universität Stuttgart, Pfaffenwaldring 57, 70569 Stuttgart, Germany
}%
\author{T. Langen}%
\affiliation{5. Physikalisches Institut and Center for Integrated Quantum Science and Technology, Universität Stuttgart, Pfaffenwaldring 57, 70569 Stuttgart, Germany
}%
\author{M. Zwierlein}%
\affiliation{MIT-Harvard Center for Ultracold Atoms, Research Laboratory of Electronics, and Department of Physics, Massachusetts Institute of Technology, Cambridge, Massachusetts 02139, USA}
\author{T. Pfau}%
\email{t.pfau@physik.uni-stuttgart.de}
\affiliation{5. Physikalisches Institut and Center for Integrated Quantum Science and Technology, Universität Stuttgart, Pfaffenwaldring 57, 70569 Stuttgart, Germany
}

\date{\today}

\begin{abstract}

We observe signatures of radial and angular roton excitations around a droplet crystallization transition in dipolar Bose-Einstein condensates. In situ measurements are used to characterize the density fluctuations near this transition. The static structure factor is extracted and used to identify the radial and angular roton excitations by their characteristic symmetries. These fluctuations peak as a function of interaction strength indicating the crystallization transition of the system. We compare our observations to a theoretically calculated excitation spectrum allowing us to connect the crystallization mechanism with the softening of the angular roton modes.

\end{abstract}

\maketitle

The roton dispersion relation is essential to understand the thermodynamics and density fluctuations in superfluid helium \cite{Landau1941,Feynman1954heTwoFluid,Feynman1953heLambdaTransition,Feynman1955heRotonCharacter}. Initially interpreted by Feynman as the ``ghost of a vanishing vortex ring" \cite{Feynman1955LowTempPhysics}, nowadays the roton is seen as a precursor to the crystallization of a system \cite{Nozieres2004}. Quantum gases with dipolar interactions feature a similar dispersion relation due to the anisotropic and long-range nature of their interaction \cite{Santos2003,Wenzel2018}. In contrast to helium, the high tunability of atomic quantum gases allows for systematic studies of roton excitations. Tuning inter-atomic interactions can soften the rotons, which trigger an instability and the formation of a crystal of quantum droplets \cite{Kadau2016,Ferrier-Barbut2016,Bottcher2020}. For elongated systems confined in cigar-shaped traps, this softening leads to the emergence of one-dimensional supersolid states that simultaneously exhibit superfluid flow and crystalline order \cite{Tanzi2019,Bottcher2019,Chomaz2019, Guo2019,Natale2019,Tanzi2019a}. 

In cylindrically symmetric oblate traps, two types of roton excitations have been predicted to play a crucial role in the instability \cite{Ronen2007,Wilson2008,Bisset2013}. These two modes are the radial and angular roton modes corresponding to the two spatial degrees of freedom in the system. The spectrum of these two-dimensional (2D) modes is more complex than the spectrum of a previously studied elongated one-dimensional (1D) supersolid \cite{Hertkorn2019}, making it a challenge to distinguish their individual contributions to the crystallization. While the roton modes were directly observed in 1D dipolar systems \cite{Chomaz2018,Petter2019,Hertkorn2020}, the various angular roton modes in 2D and their connection to the crystallization have remained elusive.

In this Letter, we observe signatures of radial and angular roton excitations of an oblate dipolar BEC around the phase transition to a 2D droplet crystal. These excitations leave their imprint in the density fluctuations of the gas, which we measure in situ, giving us direct access to the static structure factor $S(\boldsymbol{k})$ \cite{Hung2011,Hertkorn2019}. The quantity $S(\boldsymbol{k})$ features a peak at finite momentum and a distinct sixfold angular symmetry upon approaching the crystallization transition. We use mean-field simulations of the excitation spectrum to interpret the experimental results. The observed emergence of angular structure is thereby directly linked to the softening of angular roton modes.

The excitation spectrum of dipolar BECs in cylindrically symmetric traps can be theoretically studied in the Bogoliubov-de Gennes (BdG) framework. We solve the BdG-equations including the first beyond mean-field correction \cite{Ronen2006,Hertkorn2019} term known as the Lee-Huang-Yang (LHY) correction using our experimental parameters \cite{supmat}. We calculate the excitation spectrum of the low-lying modes in the BEC up to the point of the instability of $15\times 10^3$ $^{162}\mathrm{Dy}$ atoms in a trap as a function of scattering length. The trapping frequencies are ${\omega/2\pi = \left[35,\,35.1,\,110\right]\si{\hertz}}$ and the magnetic field points along $\hat{\boldsymbol{z}}$. The trap geometry is deliberately made asymmetric to numerically lift the degeneracy in the x-y-direction \cite{Gallemi2020}. The radial and angular roton modes found in these spectra can soften, by varying the scattering length, to energies much smaller than the trap frequencies \cite{Ronen2006,Ronen2007,Wilson2008,Bisset2013}.

\begin{figure}[tb!]
	\includegraphics[trim=0 0 0 0,clip,scale=0.55]{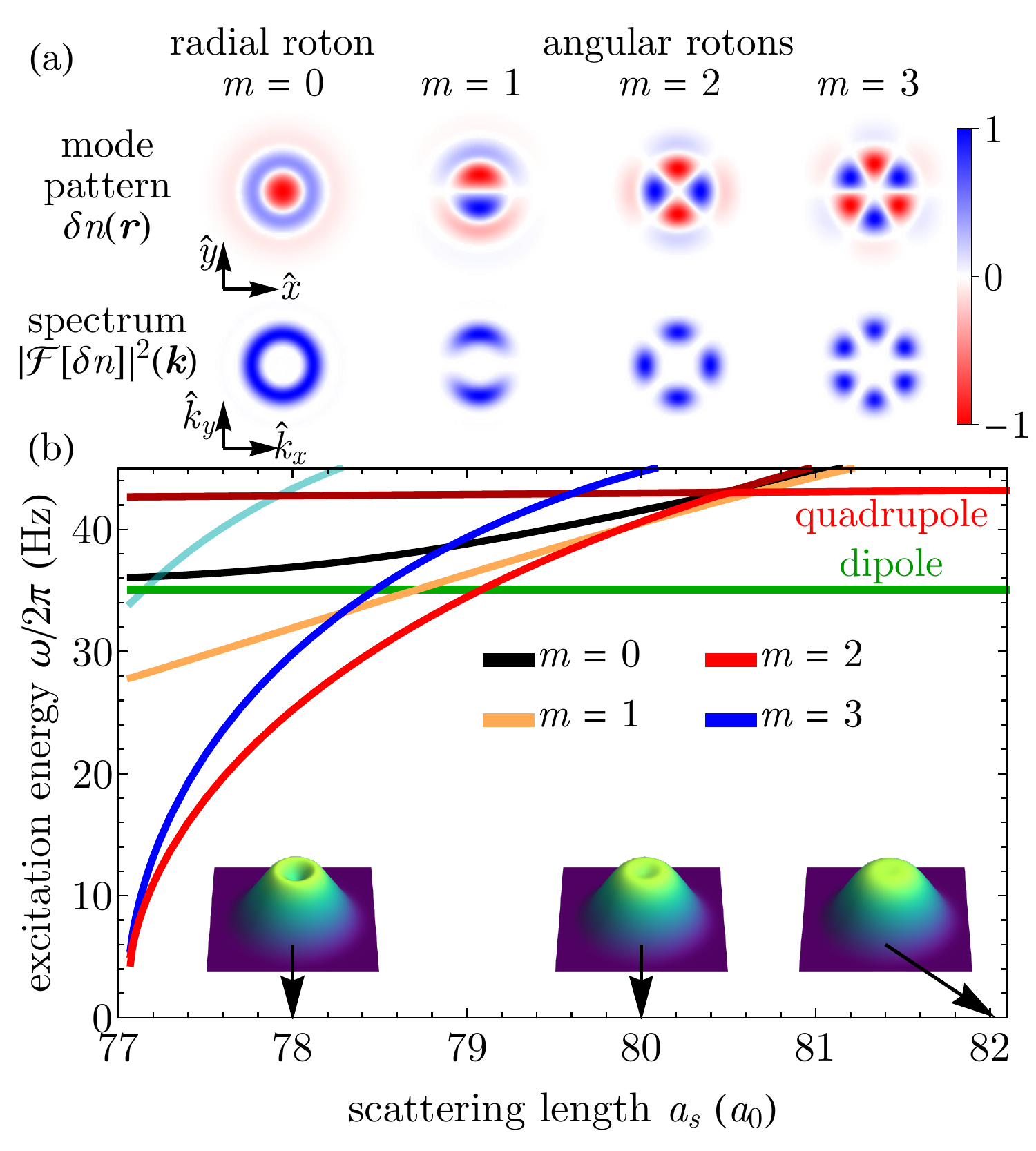}
	\caption{(a) Normalized mode patterns of the density fluctuations close to the transition at $a_s \simeq 77.5\,a_0$ and their spatial power spectra in the $x$-$y$-plane. Shown are the lowest radial $(m=0)$ and angular rotons ($m=1,\,2,\,3$), having $m$ angular nodal lines. (b) Corresponding Bogoliubov excitation energies as a function of scattering length $a_s$, with the insets indicating the emergence of a blood cell shape in the ground-state density close to the instability at $a_s \simeq 77.1\,a_0$. For decreasing scattering lengths, first the $m=2$ followed by the $m=1$ and $m=3$ angular roton modes drops below the radial dipole modes at the trap frequency and softens further towards the instability. For information on higher modes (e.g. cyan line) and details see \cite{supmat,NoteDetailsFig1}.}
	\label{fig:schematic}
\end{figure}

The mode patterns of the density fluctuation $\delta n(\boldsymbol{r})$ corresponding to the radial and angular roton modes are shown in Fig.~\ref{fig:schematic}(a). We additionally present the spatial power spectra of the mode patterns given by the squared modulus of the Fourier transform $\vert\mathcal{F}[\delta n]\vert^2(\boldsymbol{k})$, illustrating the individual contributions to the static structure factor \cite{Arkhipov2005,Astrakharchik2007,Blakie2012,Blakie2013depletionFluctuations,PitaevskiiBook2016,Hertkorn2020}. Radial roton modes are circularly symmetric and represent ring-like density modulations at non-zero radial wavevector. Angular rotons have an angular oscillatory structure in addition to the ring-like radial density modulation. We describe the angular oscillation with $\sin(m\phi)$, where $\phi$ is the azimuthal angle and the integer $m > 0$ counts the number of nodal lines in the mode pattern. The spatial power spectrum of a mode with $m$ nodal lines has a $2m$-fold symmetry resulting in a four- and sixfold symmetric power spectrum for the $m=2$ and $m=3$ angular roton modes, respectively. In contrast, the quadrupole mode and higher-lying phonon modes might feature similar azimuthal symmetries but at smaller radial wavevector.

In Fig.~\ref{fig:schematic}(b) we show the low-lying excitation energies and ground state shapes from the BEC side as a function of scattering length towards the phase transition point \footnote{The near degeneracy of the roton modes on the BEC side translates into a complex situation on the crystal side. We find a number of competing ground states reflecting the different symmetries of the rotons. This precludes a further meaningful BdG-analysis for the given parameters.}. In the BEC regime at high scattering lengths $a_s \simeq 81\,a_0$, furthest away from the transition point, the dipole mode has the lowest excitation energy and lies at exactly the trap frequency \cite{Kohn1961}. For decreasing scattering lengths, several higher-lying modes rapidly decrease in energy. These are the radial and twice degenerate angular roton modes. At around $a_s \simeq 79.1\,a_0$, the $m=2$ mode drops below the trap frequency followed by the $m=1$ and $m=3$ mode. 
Near the phase transition point at $a_s \simeq 77.1\,a_0$ the $m=2$ mode is only separated by a few Hz from the $m=3$ mode. The ordering of the angular roton modes close to the phase transition point depends on a nontrivial interplay between the trap aspect ratio and the interaction strength \cite{Wilson2009}. 

Near $a_s \simeq 80.5\,a_0$ the parabola-shaped BEC ground state transforms into a biconcave blood cell-like shape, which forms as it is energetically favorable in cylindrical geometries to push part of the density to the outer rim. This further enhance the softening of angular rotons as they become excitations of the ring-shaped region of maximal density \cite{Ronen2007}. Previous studies on blood-cell-shaped ground states \cite{Ronen2006,Ronen2007,Dutta2007,Wilson2008,Wilson2009,Wilson2009AngularCollapse,Lu2010,Blakie2012,Martin2012,Bisset2013} did not include the LHY correction that has since been shown to stabilize the system against collapse. We also find it to enlarge the parameter regime for blood-cell-shaped ground states by a range in scattering length of approximately $1\,a_0$.

We experimentally study the emergence of the angular rotons with a dipolar BEC with typically \num{15d3} $^{162}\mathrm{Dy}$ atoms at a temperature ${T\simeq \SI{20}{\nano\kelvin}}$. We adjust the crossed optical dipole trap after evaporation to an almost cylindrical trap with trapping frequencies ${\omega/2\pi = \left[35(1),37(1),110(1)\right]\si{\hertz}}$ and the magnetic field along $\hat{\boldsymbol{z}}$ \cite{supmat}. In the magnetic field range around $\SI{30}{\gauss}$ that we employ, lower three-body losses lead to droplet crystal lifetimes on the order of \SI{200}{\milli\second} after crossing the phase transition, which is an increase by a factor of ten compared to previous experiments at similar densities \cite{Bottcher2019,supmat}. We image the cloud in situ with a resolution of \SI{1}{\micro\metre} and repeat the experiment around 200 times for a statistical analysis of the atomic densities.
We further quote all scattering lengths relative to a reference scattering length $a_\mathrm{ref} = 91(10)\,a_0$ corresponding to the transition point because of an overall systematic shift in our scattering length calibration \cite{supmat}.

\begin{figure}[tb!]
	\includegraphics[trim=0mm 18mm 0 5mm,clip,scale=0.65]{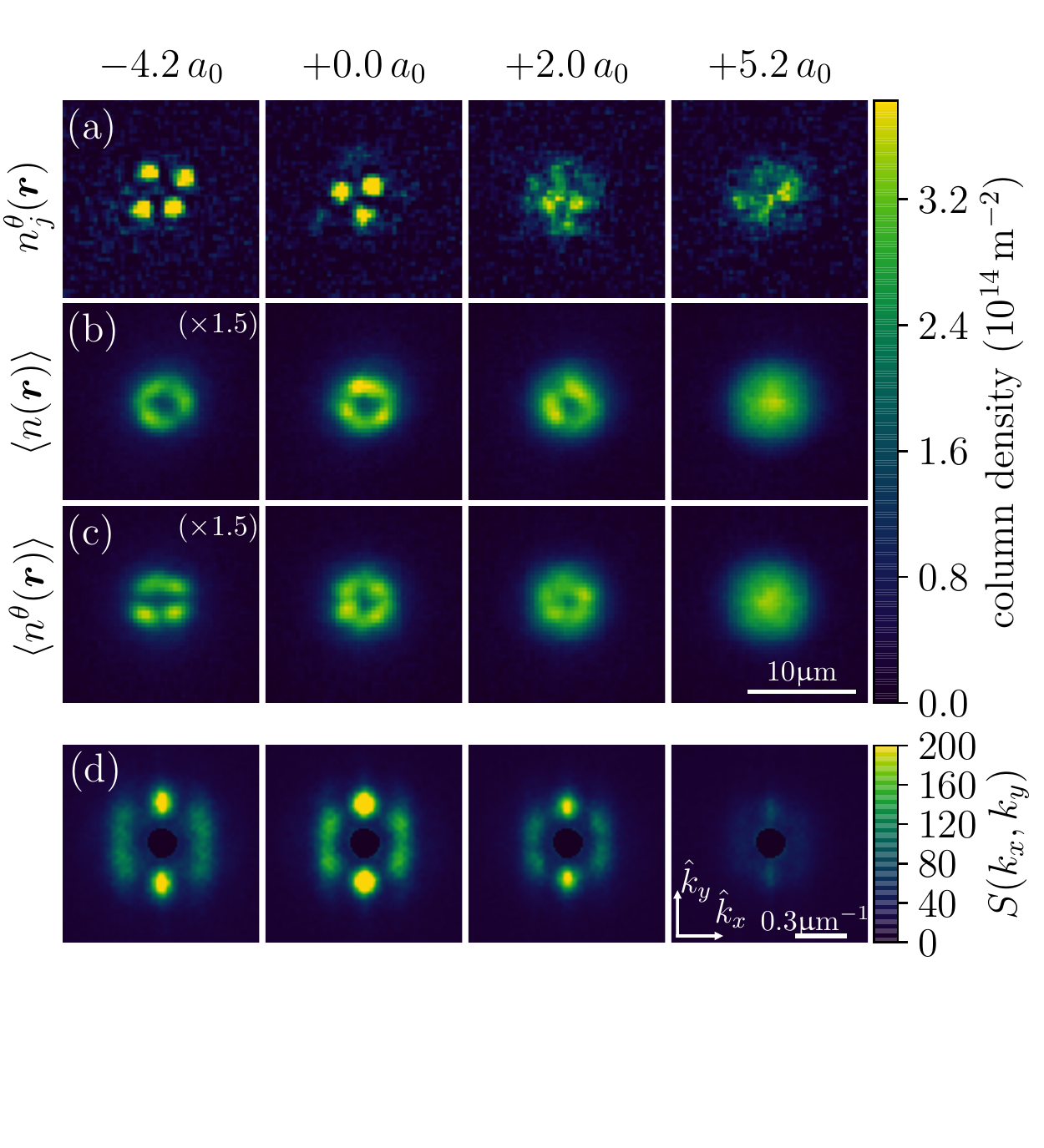}
    \caption{(a) Single-shot images for four different relative scattering lengths, in the BEC regime ($+5.2\,a_0$), closer to the transition point ($+2.0\,a_0$), in the transition region ($+0.0\,a_0$) and for a droplet crystal ($-4.2\,a_0$). (b) Mean images of the unrotated images showing no clear crystalline structure. (c) Aligned images (see text) indicate the presence of droplets in the mean image. (d) 2D structure factor showing an increasing height of the peaks at finite momentum $\vert \boldsymbol{k}\vert$ indicating the approaching transition point. The central area below $k_{\mathrm{min}}/2\pi\simeq\SI{0.11}{\micro\metre\tothe{-1}}$ (see text) was masked out.}
	\label{fig:sampleImg}
\end{figure}

We then extract the static structure factor, which connects the spectrum of elementary excitations to the major contributing modes in the density fluctuations \cite{PitaevskiiBook2016,Arkhipov2005,Astrakharchik2007,Blakie2012,Blakie2013depletionFluctuations,Hertkorn2020}. The intermediate steps of this analysis are shown in Fig.~\ref{fig:sampleImg} for four distinct scattering lengths: in the BEC regime, closer to the transition, in the transition region and for a droplet crystal.

First, we investigate the in situ densities $n_j(\boldsymbol{r})$ (Fig.~\ref{fig:sampleImg}(a)). To remove residual contributions of the dipole mode the center of mass of the atomic density distribution is taken as the origin. In addition, we post-select in an interval of $\pm 15\,\%$ with respect to the mean atom number at each scattering length \cite{supmat}. We observe that the crystal structure is randomly oriented in individual images and consequently gets washed out when averaging over many images (Fig.~\ref{fig:sampleImg}(b)). This highlights the continuous rotational symmetry breaking upon crossing the crystalline phase. In order to account for the randomly oriented crystal in the further analysis, we determine the individual rotation angles $\theta$ in Fourier space and align the single-shot images \cite{supmat}. The rotated images $n^\theta_j(\boldsymbol{r})$ are used to create new mean images ${\langle n^\theta(\boldsymbol{r})\rangle}$, which reveal the emergence of the crystal structure (Fig.~\ref{fig:sampleImg}(c)). As expected, the rotation does not affect the mean image in the BEC regime.

Second, we calculate the individual fluctuation patterns around this mean image ${\delta n^\theta_j(\boldsymbol{r})=n^\theta_j(\boldsymbol{r})-\langle n^\theta(\boldsymbol{r})\rangle}$ given by the deviations of the individual post-selected and rotated images from the mean $\langle n^\theta(\boldsymbol{r})\rangle$. We determine the mean power spectrum ${\langle \vert\delta n^\theta(\boldsymbol{k})\vert^2\rangle}$ from a Fourier transform ${\delta n^\theta_j(\boldsymbol{k})=\int\mathrm d^2r\,\delta n^\theta_j(\boldsymbol{r})e^{i\boldsymbol{k}\cdot\boldsymbol{r}}}$ of these fluctuation patterns. This mean power spectrum is closely connected to the static structure factor ${S(\boldsymbol{k})=\langle \vert\delta n^\theta(\boldsymbol{k})\vert^2\rangle/N}$ for homogeneous systems \cite{Hung2011,Hung2013,PitaevskiiBook2016} and also provides valuable insights into the nature of the excitations in non-homogeneous systems \cite{Esteve2006,Imambekov2009,Armijo2010,Jacqmin2011,Schemmer2018}. 

The resulting 2D static structure factor ${S(\boldsymbol{k})}$ is shown in Fig.~\ref{fig:sampleImg}(d). We restrict the further analysis to momenta between $k_{\mathrm{min}}/2\pi\simeq \SI{0.11}{\micro\metre\tothe{-1}}$ and $k_{\mathrm{max}}/2\pi\simeq \SI{1}{\micro\metre\tothe{-1}}$ to account for the finite size of the cloud and the finite imaging resolution in the experiment \cite{supmat}. The quantity ${S(\boldsymbol{k})}$ features several peaks that lie approximately on a ring with radius $\vert \boldsymbol{k}\vert$ around the origin. The height of the individual peaks increases when approaching the transition point. Additionally, the angular spreading of these peaks across the ring changes with scattering lengths and remains invisible in the case of unrotated images. The enhancement of the peaks along the alignment axis ($\hat{\boldsymbol{y}}$-axis) is a result of the rotation algorithm, which always aligns the images according to their individual most dominant peak in Fourier space. Exploiting the cylindrical symmetry of the trap we transform $S(\boldsymbol{k})$ to polar coordinates ${S(k_x,k_y) \to S(k,\phi)}$ and analyze its radial and angular behaviour separately. 

\begin{figure}[tb!]
	\includegraphics[trim=0 5mm 0 20mm,clip,scale=0.6]{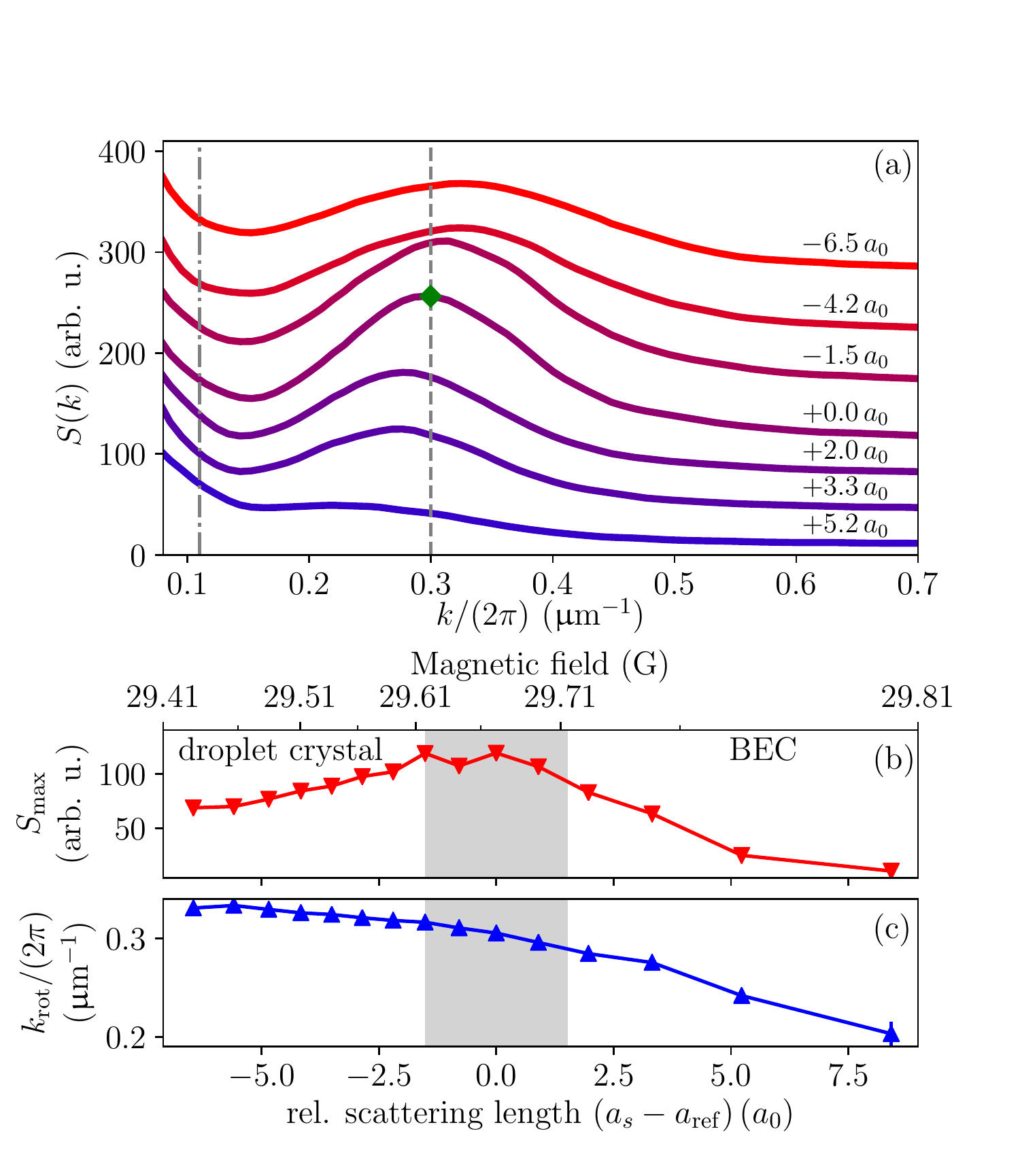}
	\caption{(a) Radial distribution $S(k)$ of the two-dimensional structure factor after integration over the angular coordinate for different relative scattering lengths. A clear peak at finite wavevector rises towards the phase transition. For clarity the lines for smaller scattering lengths were shifted vertically. (b) The amplitude of this peak is obtained from a Gaussian fit reaching its maximum at the phase transition. (c) Roton momentum shifts towards larger values in the droplet regime where it stays constant.  The dash-dotted line on the left indicates the smallest momentum $k_{\mathrm{min}}/2\pi\simeq \SI{0.11}{\micro\metre\tothe{-1}}$. The green diamond indicates the maximum $S(k)$ and the dashed line the roton momentum at that maximum. The gray area in (b) and (c) indicates the transition region.}
	\label{fig:sFactorRad}
\end{figure}

The radial behaviour $S(k)$ is analyzed by integrating over the angular direction to identify modes at finite momentum $\vert\boldsymbol{k}\vert$ independent of their angular symmetry. The result is shown in Fig.~\ref{fig:sFactorRad}(a) for scattering lengths around the transition. Starting in the BEC regime we find $S(k)$ to be relatively flat with only a small peak at around  ${k/2\pi\simeq \SI{0.22}{\micro\metre\tothe{-1}}}$. Closer to the transition this peak rises and shifts towards larger momenta. We determine the center momentum $k_\mathrm{rot}$ and amplitude $S_\mathrm{max}$ of the peak using a Gaussian fit and show it in Fig.~\ref{fig:sFactorRad}(b)-(c).

The peak of the structure factor $S_\mathrm{max}$ first increases with decreasing scattering lengths and then features a maximum before it decreases again. This is consistent with the expectation of enhanced fluctuations at the phase transition due to the softening and thermal population of several roton modes close to the phase transition \cite{Hertkorn2020}. Compared to 1D, the increased number of low-lying excitations with different symmetries results in higher shot-to-shot fluctuations in 2D. From individual images it is therefore more challenging to distinguish between an angular roton mode and a formed droplet crystal featuring the same symmetry. The formation of a droplet crystal is a direct consequence of the excitation of an angular roton mode with the same symmetry assuming a dynamical preparation scheme. We account for these uncertainties by marking a transition region ${(+0.0\pm 1.5)\,a_0}$ rather than a single point \cite{supmat}. In the droplet regime the peak amplitude decreases and its position stays approximately constant at $k_\mathrm{rot}/2\pi\simeq \SI{0.33}{\micro\metre\tothe{-1}}$ which roughly indicates the inverse droplet distance. The peak also broadens further indicating a competition between different droplet numbers and spacings.

\begin{figure}[tb!]
	\includegraphics[trim=0 5mm 0 20mm,clip,scale=0.6]{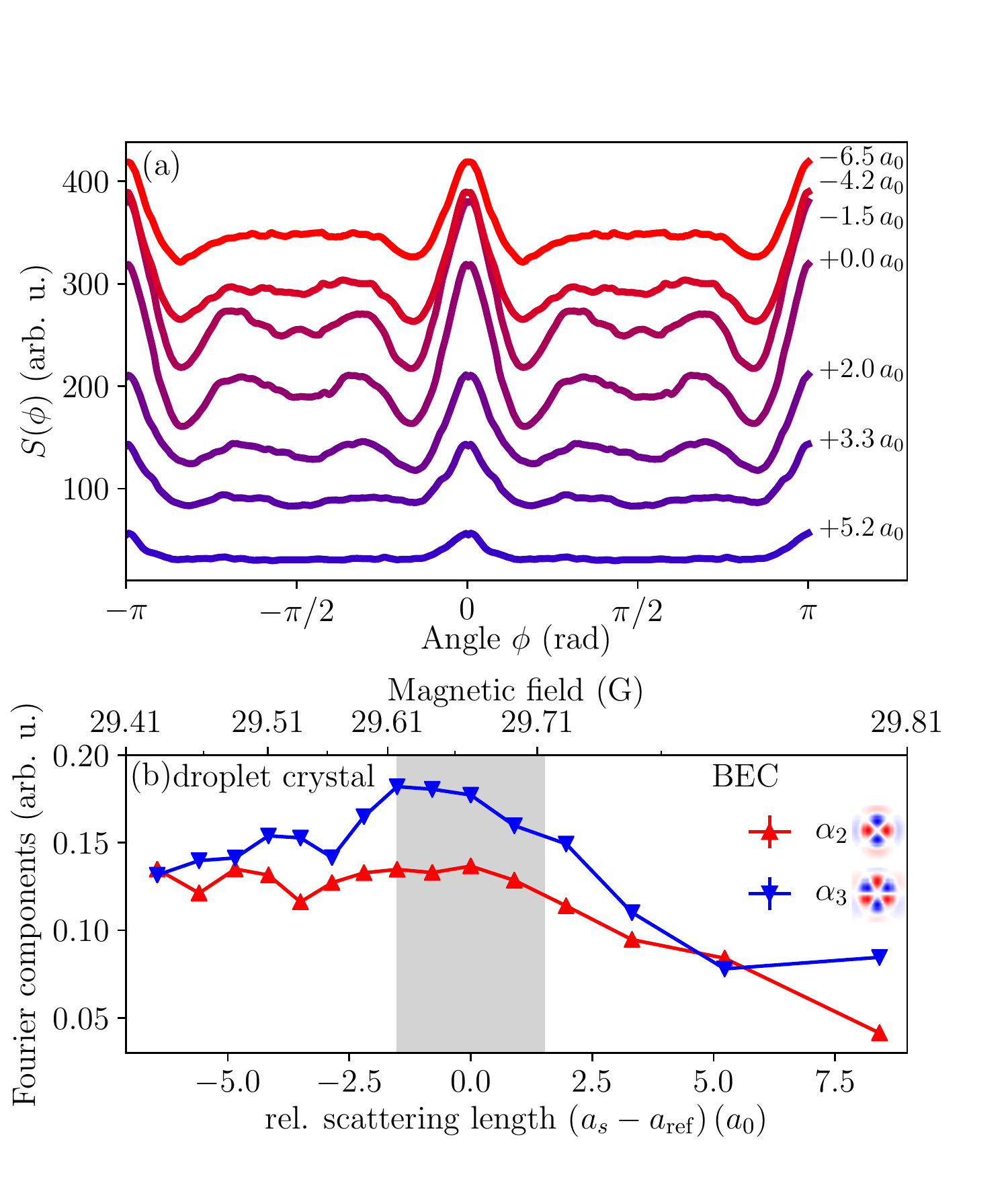}
    \caption{(a) Angular distribution $S(\phi)$ of the two-dimensional static structure factor integrated over the interval ${k/2\pi \in\left[0.2,0.45\right]\si{\micro\metre\tothe{-1}}}$ around the roton momentum $k_\mathrm{rot}$ for different relative scattering lengths. (b) Decomposition of $S(\phi)$ into Fourier components matching the symmetry of the lowest two angular roton modes $m=2$ and $m=3$ in the BEC. For clarity, the lines in (a) were shifted vertically for smaller scattering lengths. The gray area in (b) indicates the transition region.}
	\label{fig:sFactorAng}
\end{figure}

The angular behaviour $S(\phi)$ is obtained by integrating $S(k,\phi)$ over the annulus with $k\in\left[0.2,0.45\right]\si{\micro\meter\tothe{-1}}$ \cite{supmat}. $S(\phi)$ is shown in Fig.~\ref{fig:sFactorAng}(a) and allows us to attribute the enhancement of the fluctuations to individual modes. 

Starting in the BEC regime, only two comparatively small peaks at ${\phi=0}$ and ${\phi=\pi}$ are visible which increase for lower scattering lengths and can mainly be attributed to our rotation algorithm. Closer to the transition, four additional peaks at ${\phi=\pm\pi/3}$ and ${\phi=\pm2\pi/3}$ emerge to produce a rising sixfold symmetry, indicative of the $m=3$ angular roton mode. In the droplet regime these intermediate peaks start to wash out, presumably due to the competition of the three- and four-droplet configurations.

We further quantify the changing mode population with the lowest coefficients of a Fourier expansion of the periodic function  $S(\phi)$ shown in Fig.~\ref{fig:sFactorAng}(b) \cite{supmat}. On the BEC side, the Fourier coefficients $\alpha_n$ give an indication of the underlying symmetry that then can be connected to the modes from the simulation of Fig.~\ref{fig:schematic}. We focus on the coefficients $\alpha_2$ and $\alpha_3$ which both have a low contribution in the BEC regime. 

An increasing weights indicate a softening of several angular roton modes. The $\alpha_2$ and $\alpha_3$ mode increase towards the transition, in agreement with the simulation shown in Fig.~\ref{fig:schematic}(b). The $\alpha_2$ weight saturates near the transition and stays constant in the droplet regime. However, the $\alpha_3$ weight reaches a maximum after the transition and becomes smaller in the droplet regime. A stronger $\alpha_3$ weight than $\alpha_2$ is not supported by the presented theory if one assumes the population of those modes is in thermal equilibrium, as the $m=2$ mode has a slightly lower energy than the $m=3$ mode. This effect either hints towards non-equilibrium dynamics or towards the limits of the LHY approximation in the theoretical description \cite{Schutzhold2006,Lima2011,Lima2012}.

In the crystalline domain the two weights approach each other again indicating that neither of the two angular roton modes are dominant. In this regime $S(\boldsymbol{k})$ cannot be viewed as a measurement of excitations on top of the crystalline ground state because we find states with competing droplet numbers broadening the peak in the radial distribution of the 2D structure factor (see Fig.~\ref{fig:sFactorRad}(a) and \cite{supmat}). Density patterns at single scattering lengths could have two to five droplets present in each shot (see \cite{supmat}), indicating that the mean atomic density does not have same symmetry as the ground state. Then, the weights $\alpha_i$ reflect the symmetry of the observed droplet crystal rather than an excitation on top of this crystalline ground state. The similarity of the $\alpha_2$ and $\alpha_3$ weights therefore indicates similar probabilities to find a droplet crystal with fourfold or sixfold symmetry. 

In conclusion, we have reported on signatures of radial and angular roton modes by investigating the 2D static structure factor $S(\boldsymbol{k})$ of a dipolar BEC. The characteristic sixfold symmetry of $S(\boldsymbol{k})$ in the BEC regime can be identified with the population of the angular roton mode. These observations are supported by simulations of the excitation spectrum. Our study lays the foundation for a better understanding of the dominant excitations of dipolar BEC in oblate traps close to the transition to a droplet crystal. It connects the low-lying angular roton modes to the crystallization mechanism and the formation of droplets.

\begin{acknowledgements}
    This work is supported by the German Research Foundation (DFG) within FOR2247 under Pf381/16-1 and Bu2247/1, Pf381/20-1, FUGG INST41/1056-1 and the QUANT:ERA collaborative project MAQS. M.G. and M.Z. acknowledge funding from the Alexander von Humboldt Foundation.
\end{acknowledgements}

\bibliographystyle{apsrev4-1}
\bibliography{refs_cleaned} 

\begin{thebibliography}{66}
\bibitem [{Not()}]{NoteDetailsFig1}%
  \BibitemOpen
  \href@noop {} {}\bibinfo {note} {{The $m=1$ roton modes weakly lose their
  degeneracy near the transition due to the slightly asymmetric trap used in
  the simulations. We show the average of the two energies in
  Fig.~\ref{fig:schematic} and refer to \cite{supmat} for further
  details.}}\BibitemShut {Stop}%
\end{thebibliography}

\clearpage

\section{Supplemental Material}

\setcounter{figure}{0}

\renewcommand{\figurename}{Supplementary Figure}
\renewcommand{\thefigure}{S\arabic{figure}} 

\subsection{Experimental protocol and Feshbach resonances}

The complete experimental procedure has been described in detail in Refs \cite{Kadau2016,Bottcher2019,Guo2019}. We prepare a quasipure BEC of $^{162}$Dy with ${T\simeq 20\,\si{\nano\kelvin}}$ in a crossed optical dipole trap at \SI{1064}{\nano\meter}. We then adiabatically reshape the trap to trapping frequencies ${\omega / 2 \pi = \lbrack 35(1),\,37(1),\,110(1)\rbrack\,\si{\hertz}}$ within \SI{200}{\milli\second} and subsequently ramp the magnetic field in \SI{50}{\milli\second} from the evaporation field (\SI{30}{\gauss} to between  \SI{29.430(1)}{\gauss} and \SI{29.805(1)}{\gauss}. The magnetic field is always oriented along the $\hat{\boldsymbol{z}}$-direction. The magnetic field difference corresponds to a change in scattering length of about $14.8(5)\,a_0$. This long ramp time and a subsequent wait of additional \SI{80}{\milli\second} ensures that the system has enough time to equilibrate as much as possible before the atomic cloud is imaged in situ with a microscope objective with a resolution of about \SI{1}{\micro\metre}.

We use two broader Feshbach resonances within the rich spectrum of resonances in $^{162}\mathrm{Dy}$ \cite{Baumann2014,Maier2015,Maier2015a,Frisch2014}. We determine the position of these resonances to be at ${B_0=\SI{22.5(5)}{\gauss}}$ and ${B_1=\SI{26.7(5)}{\gauss}}$ with widths of ${\Delta_0=\SI{2.3(5)}{\gauss}}$ and ${\Delta_1=\SI{100(10)}{\milli\gauss}}$, respectively. The characterization of the broad resonance is challenging as it overlaps with a dense group of narrow resonances \cite{Lucioni2018}. These two resonances dominate the scattering length calibration and allow us to work several Gauss away to significantly reduce three-body losses. The lifetime of droplets several $a_0$ away from the phase transition is about \SI{200}{\milli\second} in the used range of magnetic fields, which is longer than in our previous measurements \cite{Bottcher2019}. The scattering length calibration is perturbed further by two small resonances near \SI{30}{\gauss}, the magnetic field we work at. These resonances are located at ${B_2=\SI{29.341(2)}{\gauss}}$ and ${B_3=\SI{29.860(2)}{\gauss}}$ with widths of ${\Delta_2=\SI{3(1)}{\milli\gauss}}$ and ${\Delta_3=\SI{1(1)}{\milli\gauss}}$, respectively. The relatively large error in the width of the broad resonances comes from these narrow resonances being near the zero-crossing of the broad resonance. By changing the width of the broad resonance within the stated uncertainty the scattering length calibration can easily be shifted by several $a_0$. We work far from the resonances so that this uncertainty only leads to an overall shift in the scattering length. Therefore we can determine differences in scattering length with a higher precision than the overall shift. We quote all scattering lengths throughout this work relative to a reference scattering length ${a_\mathrm{ref} = 91(10)\,a_0}$ corresponding to the center of the transition region. The error includes the mentioned uncertainty in the width of the broad Feshbach resonances. The background scattering length ${a_\mathrm{bg} = 140(20)\,a_0}$ also suffers from a large uncertainty \cite{Tang2015a, Tang2016, Tang2018}.

\begin{figure}[tb!]
	\includegraphics[trim=0 0 0 0,clip,scale=0.7]{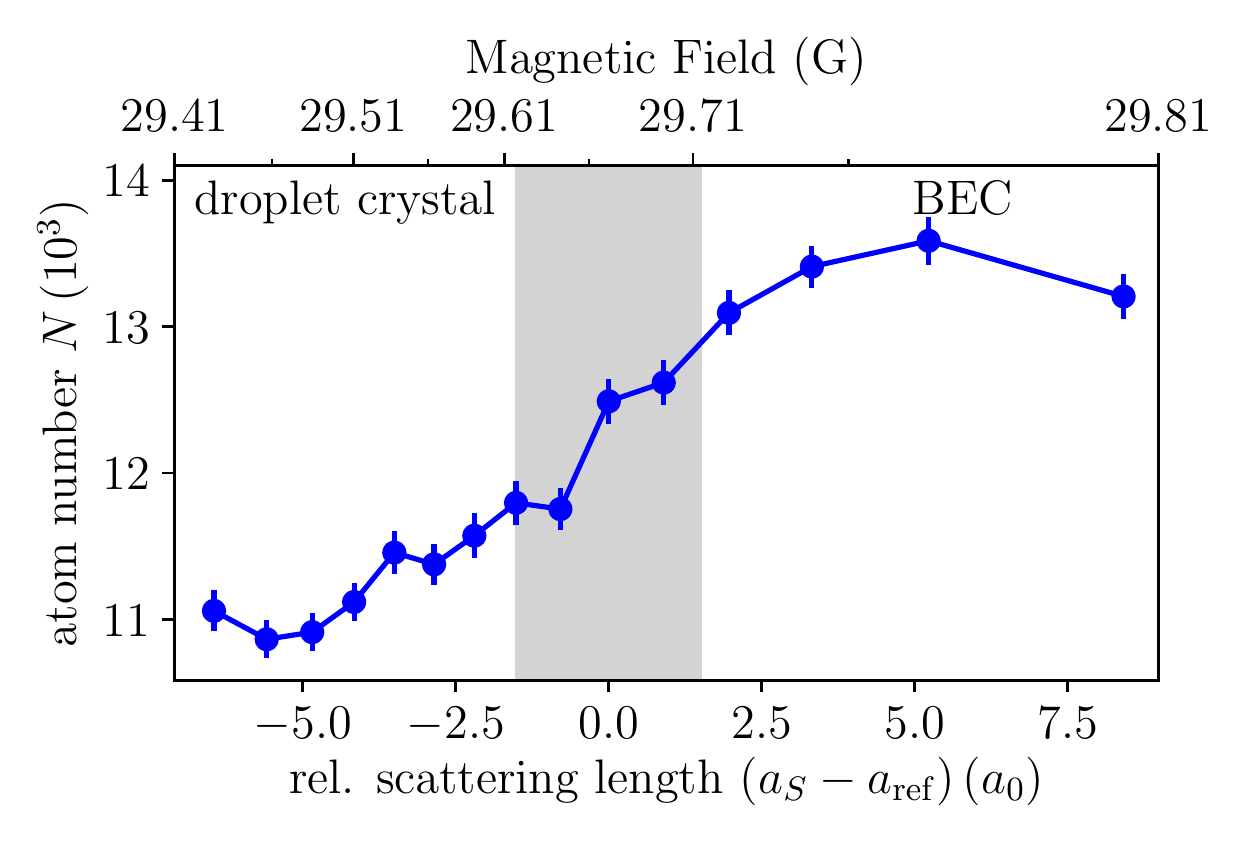}
	\caption{Average atom number for the scattering length range in the experiment. Crossing the phase transition from BEC to droplet crystal the increasing density leads to larger three-body losses causing lower atom numbers at smaller scattering lengths. Error bars indicate the standard error of the mean.}
	\label{fig:atomnumber}
\end{figure}

For our configuration, lower magnetic fields correspond to lower scattering length. The densities and three-body losses increase at lower scattering lengths, leading to a lower mean atom number in the droplet. Figure~\ref{fig:atomnumber} shows the mean atom number for the complete scattering length range. The mean atom number is about $20\,\%$ lower in the droplet regime than the BEC regime.

\subsection{Analysis method and angular-averaged static structure factor}

The procedure to extract the static structure factor from the individual in situ image is similar to the one described in \cite{Hertkorn2020}. After centering the images to their center-of-mass to remove the effect of the dipole mode, we perform a post-selection on atom number for every scattering length. We only consider images with atom numbers of $\pm 15\,\%$ around the mean atom number at every scattering length.

The rotational symmetry of the trap generates rotationally symmetric averaged density distributions that washes out the droplet structure found in individual images. We therefore remove the rotational symmetry and align the droplet crystals from each experimental realization to the same angle. After calculating the 2D Fourier transform we transform the image to polar coordinates and integrate over radial momenta larger than the small-$k$ threshold $k>k_\mathrm{min}$. We then find the angle $\theta$ that corresponds to the maximum and rotate the original image by $-\theta$. Minor imperfections in the imaging system such as astigmatism and distortion can lead to a non-uniform distribution of the rotation angles. This effect becomes stronger in the BEC where there is no strong feature at finite momentum. These arbitrary rotations should also not affect the evaluation because there is no angular structure.

The calculation of the static structure factor contains an ensemble average of the power spectrum of the fluctuations which is sensitive to single-shot atom number fluctuations. We reduce this effect by normalizing each rotated image to its specific atom number $\tilde{n}^\theta_j = n^\theta_j/N_j$ and calculate the normalized fluctuations $\delta \tilde{n}^\theta_j(\boldsymbol{r}) = \tilde{n}^\theta_j(\boldsymbol{r}) - \braket{\tilde{n}^\theta(\boldsymbol{r})}$ as the deviation of the normalized image $\tilde{n}^\theta_j(\boldsymbol{r})$ from the mean normalized image $\braket{\tilde{n}^\theta(\boldsymbol{r})}$. The structure factor is given by
\begin{equation}\label{eq:Sfac}
S(\boldsymbol{k}) = \bar{N}\braket{ | \delta \tilde{n}^\theta(\boldsymbol{k}) |^2 },
\end{equation}
where $\braket{ \vert \delta \tilde{n}^\theta(\boldsymbol{k})\vert^2 }$ is the mean power spectrum of these fluctuations and $\bar{N}$ is the mean atom number. The 2D Fourier transform of the normalized fluctuations $\delta \tilde{n}^\theta_j (\boldsymbol{k}) = \mathcal{F}[\delta \tilde{n}^\theta_j] (\boldsymbol{k}) = \int\!\mathrm{d}^2r\,\delta \tilde{n}^\theta_j (\boldsymbol{r})e^{i \boldsymbol{k}\cdot \boldsymbol{r}}$ is obtained from the line integrated images. The presented measurement of the static structure factor is limited at small-$k$ and at large-$k$. The finite size of the atom cloud $L\simeq 9\,\si{\micro\meter}$ sets the small-$k$ limit at ${k_{\mathrm{min}}/2\pi \simeq  0.11\,\si{\micro\meter\tothe{-1}}}$. The large-$k$ limit is set by the finite imaging resolution of \SI{1}{\micro\metre} or ${k_{\mathrm{max}}/2\pi \simeq  1\,\si{\micro\meter\tothe{-1}}}$.

We transform the two-dimensional static structure factor to polar coordinates $S(k_x,k_y) \to S(k,\phi)$ and then analyse the radial and angular dependence separately. The angular distribution of the static structure factor $S(\phi)$ is a periodic function by definition, and it is possible to write it as a series expansion in harmonic function ${f(x)=\sum_n \alpha_n\cos(nx+\varphi_n)}$ with $x\in \left[0,2\pi\right)$. The coefficients ${\alpha_n=\sqrt{a_n^2+b_n^2}}$ and the phase ${\varphi_n=\arctan(a_n/b_n)}$ are given by ${a_n=(1/\pi)\int_{0}^{2\pi}f(x)\cos(nx)\mathrm{d}x}$ and ${b_n=(1/\pi)\int_{0}^{2\pi}f(x)\sin(nx)\mathrm{d}x}$ respectively. The weights $\alpha_i$ can be related to the weight of the corresponding mode in the Bogoliubov-de Gennes calculation because these harmonic functions have exactly the same symmetry as the density fluctuation patterns of the various angular roton modes. The angular structure factor was normalized to its maximum value before calculating the Fourier components, minimizing the influence of the absolute scale at each scattering length.

\subsection{Determination of the phase transition region}

\begin{figure}
    \includegraphics[trim=0 0 0 0,clip, scale=0.7]{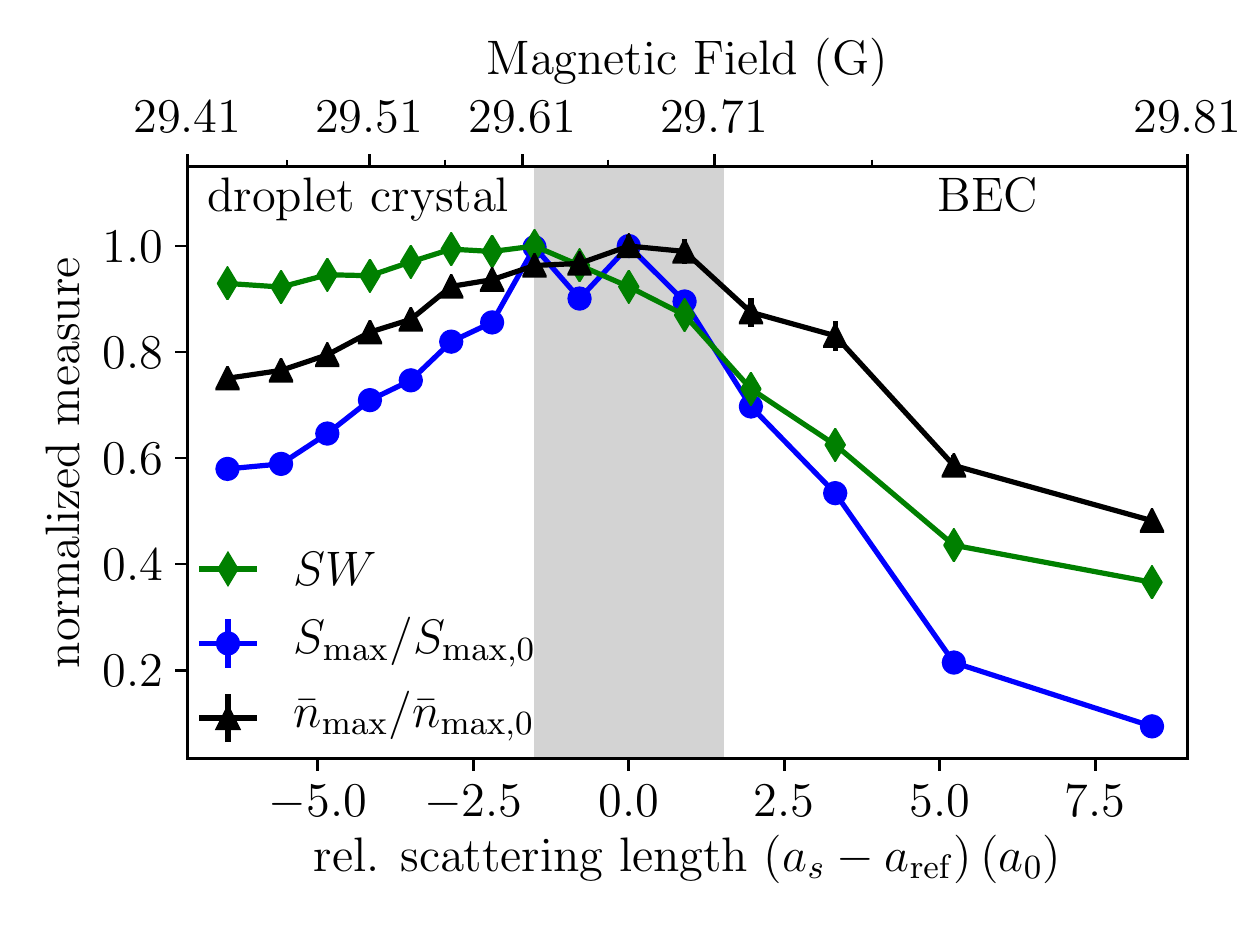}
    \caption{Three different measures to characterize the phase transition are shown. The peak amplitude of the structure factor $S_\mathrm{max}$ (blue), the spectral weight $SW$ (green), and the average peak density $n_\mathrm{max}$ (black) are all normalized to their maxima. While the $SW$ and $S_\mathrm{max}$ increase around $+1.5\,a_0$, the peak density reaches values close to its maximum at higher relative scattering lengths. The shaded gray area indicates the region where the phase transition occurs and was determined by these measures (see text). The maximum values are $\bar{n}_\mathrm{max,0}=\SI{7.2e14}{\meter\tothe{-2}}$ and $S_\mathrm{max,0}=\num{119}$.}
    \label{fig:supmatPhaseTrans}
\end{figure}

The transition from a BEC to a droplet crystal occurs approximately where the peak of the static structure factor at finite momentum $S_\mathrm{max}$ reaches a maximum as a function of scattering length (see Fig.~\ref{fig:sFactorRad} of the main text). We extract this amplitude $S_\mathrm{max}$ by fitting a $S(k)$ using a Gaussian fit after subtracting the background. In Fig.~\ref{fig:supmatPhaseTrans} we compare $S_\mathrm{max}$ with two other measures to define the phase transition: the average peak density $\bar{n}_\mathrm{max}$ and the average spectral weight $SW$. The spectral weight is defined as the integral of the spatial power spectrum $\vert n(k) \vert^2$ over the annulus with $k\in\left[0.2,0.45\right]\si{\micro\meter\tothe{-1}}$. We normalize all quantities to their maximum for comparison.

Starting in the BEC both the spectral weight $SW$ and the peak amplitude of the structure factor $S_\mathrm{max}$ show a strong increase towards lower scattering lengths. $S_\mathrm{max}$ seems to have a extended peak between $-1.5\,a_0$ and $+0.0\,a_0$ and decreases again for even lower scattering length. The spectral weight $SW$ saturates around $+0.0\,a_0$, indicating the transition to a droplet crystal. By contrast, the peak density increases at higher scattering lengths in the BEC and reaches its maximum at $+1.0\,a_0$. 

The maximum peak amplitude in the static structure factor hints towards enhanced fluctuations close to the transition point due to a softening and thermal population of multiple roton modes. The peak density and the spectral weight are useful indicators of this transition but they cannot distinguish between an angular roton mode of a certain symmetry and a formed droplet crystal with the same symmetry. The increased number of low-lying modes with different symmetries in oblate trap geometries might lead to a higher shot-to-shot fluctuations. These enhanced fluctuations might result in small differences between the three measures. We therefore identify a transition region centered at $+0.0\,a_0$ with a width of $\pm 1.5\,a_0$ (shaded gray area).

\subsection{Droplet number probability distribution}

\begin{figure}[tb!]
	\includegraphics[trim=0 0 0 0,clip,scale=0.7]{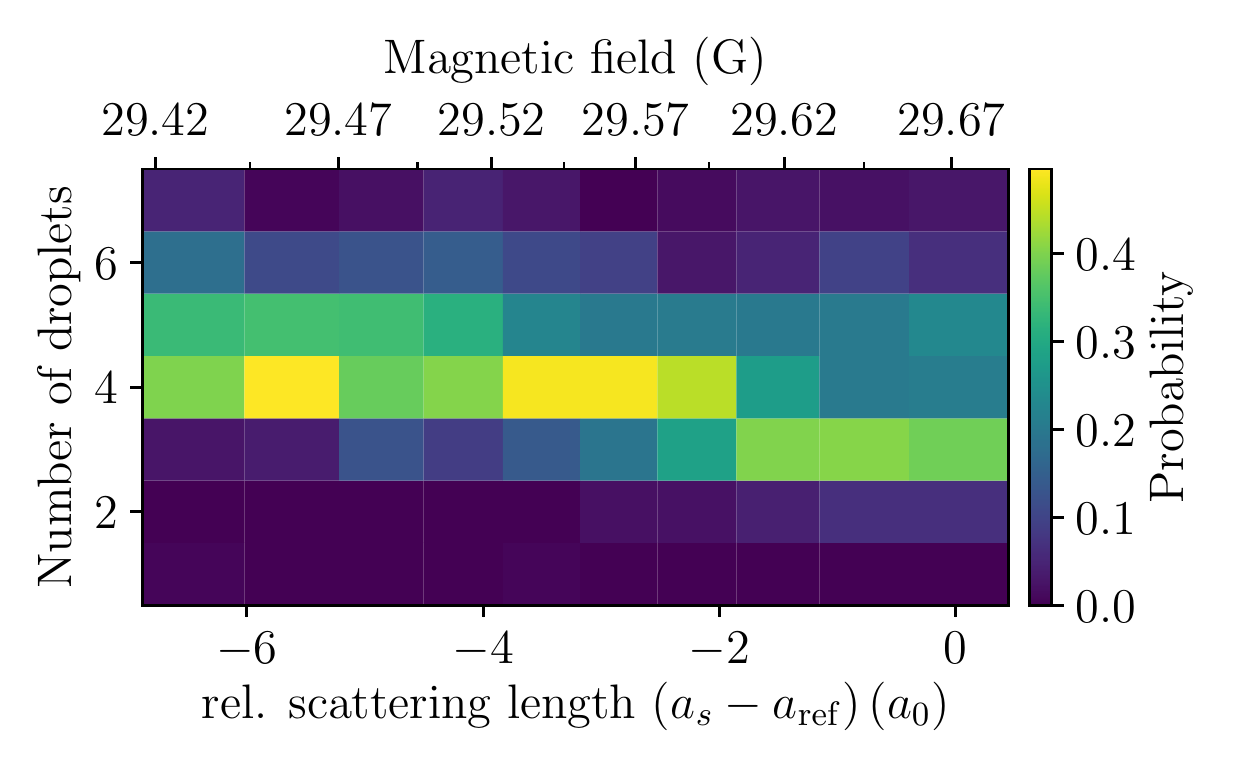}
	\caption{Droplet number as a function of the rel. scattering length in the droplet regime. Starting with a maximum probability for three droplets right after the phase transition, the peak of the distribution shifts towards four droplets for lower scattering length.}
	\label{fig:supmatDropNum}
\end{figure}

The angular static structure factor $S(\phi)$ for scattering lengths in the droplet regime suggests the competition between a fourfold and sixfold symmetric droplet crystal. Both symmetries can be seen in the Fourier component analysis in Fig.~\ref{fig:sFactorAng}(b) of the main text. To identify the two symmetries, we determine the number of droplets per image and compare the droplet number probability distribution for different scattering lengths. We compute the number of droplets per image by using a droplet number detection algorithm based on the Laplacian-of-Gaussian technique \cite{Lindeberg1993,Lindeberg1998}. The resulting histograms for several scattering lengths in the droplet regime are shown in Fig.~\ref{fig:supmatDropNum} as a color plot.

We are most likely to observe three droplets directly after entering the droplet regime, but four droplets are more probable deeper in the droplet regime. This change in the average droplet number indicates a change of the droplet crystal lattice structure. The lattice structure is a triangle for three droplets and a square for four droplets. The varying probability of droplet number indicates the competition between the sixfold (three droplets) and fourfold (four droplets) symmetry in the static structure factor.

\subsection{Simulation details and higher-lying excitation modes}

We describe our system of dipolar atoms in BEC close to the crystallization transition by the extended Gross-Pitaevskii equation (eGPE)  \cite{Ronen2006,Wenzel2017,Roccuzzo2019} and refer to Ref.~\cite{Hertkorn2019} for details on the simulations.

\begin{equation}\label{eq:GPE}
i \hbar \partial_t \psi = H_\mathrm{GP} \psi,
\end{equation}
where we define ${H_\mathrm{GP} = H_0 + g|\psi|^2 + \Phi_\mathrm{dip}[\psi] + g_\mathrm{qf} |\psi|^3}$ and $\psi$ is normalized to the atom number ${N=\int \mathrm{d}^3r\, |\psi(\boldsymbol{r})|^2}$. The term ${H_0 = -\hbar^2 \nabla^2 / 2M + V_\mathrm{ext}}$ contains the kinetic energy and trap confinement ${V_\mathrm{ext}(\boldsymbol{r}) = M(\omega_x^2 x^2 + \omega_y^2 y^2 + \omega_z^2 z^2)/2}$, where $M$ is the mass.  The contact interaction strength $g = 4\pi\hbar^2a_s/M$ is given by the scattering length $a_s$. The dipolar mean field potential is $\Phi_\mathrm{dip} = \int \mathrm{d}^3r'\,  	V_\mathrm{dd}(\boldsymbol{r}-\boldsymbol{r}') |\psi(\boldsymbol{r}', t)|^2$ where $V_\mathrm{dd}(\boldsymbol{r}) = \frac{3g_\mathrm{dd}}{4\pi} \frac{1-3\cos^2 \vartheta}{|\boldsymbol{r}|^3}$ is the dipolar interaction for aligned dipoles with an angle $\vartheta$ between $\boldsymbol{r}$ and the magnetic field axis. The strength of the dipolar interaction is given by the parameter $g_\mathrm{dd} = 4\pi\hbar^2a_\mathrm{dd}/M$ characterized by the dipolar length $a_\mathrm{dd} = \mu_0 \mu_m^2 M / (12 \pi \hbar^2)$ and the magnetic moment $\mu_m$. Furthermore the quantum fluctuations within the local density approximation are given by the quantity ${g_\mathrm{qf} = 32/(3\sqrt{\pi}) g a^{3/2} Q_5 (\epsilon_\mathrm{dd})}$, where $\epsilon_\mathrm{dd} = g_\mathrm{dd} / g = a_\mathrm{dd} / a_s$ is the relative dipolar strength. In our simulations, we use the approximation  ${Q_5(\epsilon_\mathrm{dd}) \simeq 1+ \frac{3}{2}\epsilon_\mathrm{dd}^2}$ \cite{Lima2012,Ferrier-Barbut2016,Wenzel2017,Bisset2016,Baillie2017}.

\begin{figure}[tb!]
	\includegraphics[trim=0 0 0 0,clip,scale=0.55]{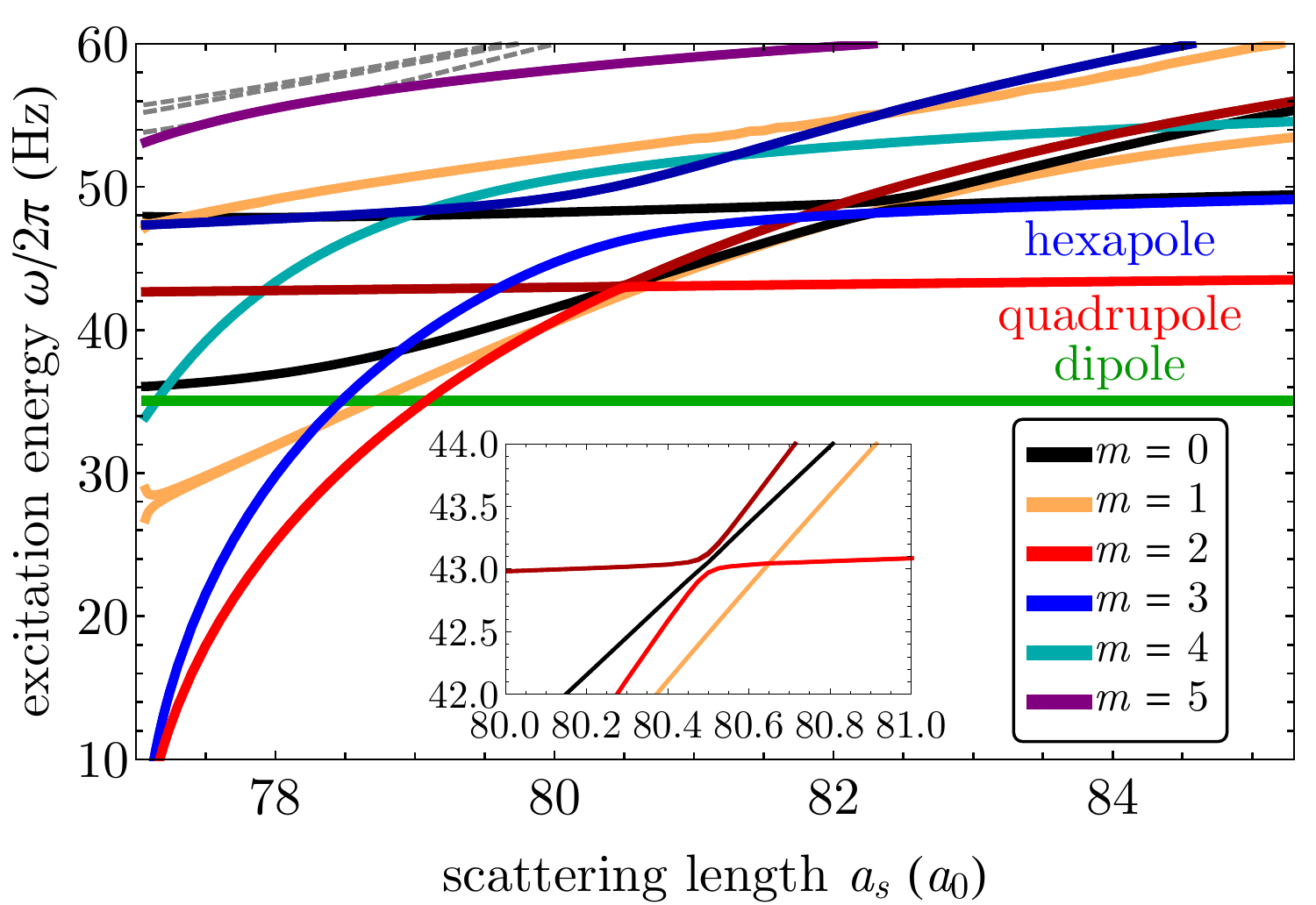}
	\caption{The excitation spectrum shown in Fig.~\ref{fig:schematic} of the main text for higher excitation energies and larger scattering lengths. The quasi-periodic boundary condition along the angular direction results in multiple angular roton modes that all soften towards the transition point. The first five are shown here including their coupling to the dipole and quadrupole mode of the BEC. As an inset we show the avoided crossing of the $m=2$ angular roton mode with the quadrupole mode in greater detail.}
	\label{fig:supmatSpectrum}
\end{figure}

We use the Bogoliubov-de Gennes (BdG) formalism as described in Ref.~\cite{Hertkorn2019} and linearly expand the wavefunction ${\psi(\boldsymbol{r},t) = \psi_0(\boldsymbol{r}) + \lambda \lbrack u(\boldsymbol{r}) e ^{-i\omega t} + v^{\ast}(\boldsymbol{r})e^{i\omega t}\rbrack e^{-i\mu t/\hbar}}$ around the ground state $\psi_0$ with the chemical potential $\mu$. This ansatz together with equation~\eqref{eq:GPE} leads to a system of linear equations that can be expressed in matrix form. See Refs. \cite{Baillie2017,Chomaz2018,Hertkorn2019} for the complete form of the BdG matrix representation. We numerically solve these equations to obtain the modes $u$ and $v$ corresponding to the lowest excitation energies $\hbar \omega$. From the solutions $u$ and $v$ we obtain the density fluctuation patterns $\delta n \propto (u+v)\psi_0$ \cite{Ronen2006,Hertkorn2019} discussed in the main text (see Fig.~\ref{fig:schematic}). We obtain a discrete spectrum of elementary excitations because of the finite-sized system.

In Fig.~\ref{fig:supmatSpectrum} we show a wider view of the excitation spectrum discussed in the main text (Fig.~\ref{fig:schematic}) with higher excitation energies and larger scattering lengths. The shown excitation spectrums of a 2D and a 1D BEC are similar \cite{Hertkorn2019}, but the inclusion of more modes complicates the 2D BEC spectrum. Due to the quasi-periodic boundary conditions along the angular direction, multiple (angular) roton modes exist, which all soften near the transition point. Which mode softens  fastest is determined by an interplay between the interaction strengths and the aspect ratio \cite{Wilson2009}. 

Generally, modes in the same $m$-subspace can couple to each other \cite{Neumann-Wigner1929,landau1981quantum,Ronen2006,Wilson2009,Bisset2013}. While the angular roton modes are softening, they couple to the BEC phonon modes in the same $m$-subspace. This can lead to avoided level crossings and hybridization between angular roton modes and BEC multipole (dipole, quadrupole, hexapole, ...) modes \cite{Bisset2013}. As an example we show an inset of the avoided crossing between the $m=2$ angular roton mode and the quadrupole mode of the BEC. The hybridization between the angular roton mode and the quadrupole mode leads to an additional radial node in the mode pattern of the quadrupole mode after the anticrossing ($a_s \lesssim 80.5$). Similar patterns can also be seen for higher angular roton modes. The radial roton mode ($m=0$) shows such an avoided crossing with the breathing mode at around $a_s \simeq 82.4\,a_0$ and the $m=3$ angular roton mode with the hexapole mode at around $a_s \simeq 80\,a_0$. Here, the deformation of the branches is larger compared to the avoided crossing of the $m=2$ angular roton mode, suggesting a larger coupling between the two participating modes.

We note that labelling the angular roton modes with their number of nodal lines $m$ is only a good choice for a circularly symmetric ground state. With an increasing asymmetry in the radial trap aspect ratio this degeneracy is lifted and $m$ ceases to uniquely classify the mode patterns. The splitting of the $m=1$ mode close to the instability is a first signature of this effect, even though the deviation from a perfectly circular trap is about $0.3\,\%$.

\end{document}